# The thermodynamic meaning of local temperature of nonequilibrium open quantum systems

LvZhou Ye,[1] Xiao Zheng,[1, 2, *] YiJing Yan,[1, 3] and Massimiliano Di Ventra[4, †]

[1] *Hefei National Laboratory for Physical Sciences at the Microscale, University of Science and Technology of China, Hefei, Anhui 230026, China*
[2] *Synergetic Innovation Center of Quantum Information and Quantum Physics, CAS Center for Excellence in Nanoscience, University of Science and Technology of China, Hefei, Anhui 230026, China*
[3] *iChEM (Collaborative Innovation Center of Chemistry for Energy Materials), University of Science and Technology of China, Hefei, Anhui 230026, China*
[4] *Department of Physics, University of California, San Diego, La Jolla, California 92093*

Measuring the local temperature of nanoscale systems out of equilibrium has emerged as a new tool to study local heating effects and other local thermal properties of systems driven by external fields. Although various experimental protocols and theoretical definitions have been proposed to determine the local temperature, the thermodynamic meaning of the measured or defined quantities remains unclear. By performing analytical and numerical analysis of bias-driven quantum dot systems both in the noninteracting and strongly-correlated regimes, we elucidate the underlying physical meaning of local temperature as determined by two definitions: the zero-current condition that is widely used but not measurable, and the minimal-perturbation condition that is experimentally realizable. We show that, unlike the zero-current one, the local temperature determined by the minimal-perturbation protocol establishes a quantitative correspondence between the nonequilibrium system of interest and a reference *equilibrium* system, provided the probed system observable and the related electronic excitations are fully local. The quantitative correspondence thus allows the well-established thermodynamic concept to be extended to nonequilibrium situations.

Probing the variation of local temperatures in systems out of equilibrium has become a subject of intense experimental interest in physics [1–5], chemistry [6–8] and life sciences [9–12]. With the development of high-resolution thermometry techniques, measurement of some sort of temperature distributions of nonequilibrium systems has been realized, such as in graphene-metal contacts [4], gold interconnect structures [5], and living cells [12].

Local electronic and phononic excitations occur in nanoelectronic devices subject to a bias voltage or thermal gradient, and hence the devices are supposedly at a local temperature somewhat higher than the background temperature. Such local heating affects crucially the device properties [13–16], and have significant influence on some physical processes, such as thermoelectric conversion [17–19], heat dissipation [8, 19], and electron-phonon interactions [20, 21]. All these studies, however, leave open the question of what precisely is a "local temperature" in a nonequilibrium system, a concept that has a well-established meaning only in global equilibrium.

Over the past decade, numerous experimental [15, 22–27] and theoretical [28–38] efforts have been made to provide practical and meaningful definitions of local temperature for nonequilibrium systems that bear a close conceptual resemblance to the thermodynamic one. However, it has remained largely unclear how to physically interpret the defined local temperature, and how to associate the measured value with the magnitudes of local excitations and local heating at a quantitative level.

This work aims at elucidating these fundamental issues through analytical and numerical analysis on nonequilibrium quantum dot (QD) systems. In particular, we shall focus on the definition of local temperature based on the zero-current condition (ZCC) proposed by Engquist and Anderson [39], and that based on the minimal-perturbation condition (MPC) as proposed in Refs. [30, 37]. Here, we will focus only on the electronic contribution to the local temperature and leave to future studies the analysis of the effects of phonons.

*Two different definitions of local temperature.* The definition of local temperature based on the ZCC has been used extensively in the literature [29, 32, 34, 35, 40–44]. The basic idea is to couple an ideal potentiometer/thermometer (the probe) to the nonequilibrium system of interest. By varying the temperature ($T_p$) and chemical potential ($\mu_p$) of the probe until *both* the electric current and heat current flowing through the probe vanish, the local temperature $T^*$ and local chemical potential $\mu^*$ of the system are determined as $T^* = T_p$ and $\mu^* = \mu_p$, respectively [45]. The ZCC is often referred to as the "local equilibrium condition", as the determined local temperature can be understood from the perspective of the zeroth law of thermodynamics. However, such a *macroscopic* definition of local temperature does not reflect the *microscopic* change of the system state in a nonequilibrium situation. Moreover, it is important to note that, unlike charge currents, we have no means to *directly* measure heat currents, since in the latter case we have no equivalent apparatus like the ammeter in the electronic case [16]. This is an often ignored but very important issue that severely limits the experimental application of the ZCC-based definition.

The MPC-based definition is conceptually different (see Fig. 1 for a schematic). Consider, for instance, a QD connected to two leads ($L$ and $R$), with the lead temperatures (chemical potentials) being $T_L$ and $T_R$ ($\mu_L$ and $\mu_R$), respectively. By locally coupling a probe ($p$) to the QD, the expectation value of a local observable $O = \langle \hat{O} \rangle$ is subjected to a perturbation that depends explicitly on $T_p$ and $\mu_p$. Within the MPC, the local temperature $T^*$ of the QD is determined by varying $T_p$ so that the perturbation $\delta O_p(T_p, \mu_p)$ is minimized [30, 37, 45]:

$$T^* = \arg\min_{T_p} |\delta O_p(T_p, \mu_p = \mu^*)| \,. \tag{1}$$

During the variation of $T_p$, $\mu_p$ is kept aligned with a preset local chemical potential $\mu^* = \zeta_L \mu_L + \zeta_R \mu_R$, where coefficients $\zeta_L$ and $\zeta_R$ are determined by measuring the electric currents [45].

At zero bias $T^*$ determined by Eq. (1) recovers exactly the physical equilibrium temperature. While in many cases the MPC-defined $T^*$ is numerically close to that obtained by the ZCC [37], the former does not require the measurement of heat currents, and hence its experimental realization is feasible. Despite this added benefit, it remains unclear how $T^*$ determined by Eq. (1) is quantitatively related to the electronic excitations in a nonequilibrium system, and what is the underlying origin of the difference between the ZCC and MPC based definitions. This leads us to question to what extent can we assign to the MPC quantity the meaning of a "thermodynamic temperature" as in the equilibrium case.

To address these fundamental issues, we consider a QD described by the single-impurity Anderson model [46]. The dot Hamiltonian is $\hat{H}_{\rm dot} = \epsilon_d\,(\hat{n}_\uparrow + \hat{n}_\downarrow) + U\hat{n}_\uparrow \hat{n}_\downarrow$, where $\hat{n}_s$ is the occupation number operator for spin–$s$ electrons, $\epsilon_d$ the dot level energy, and $U$ the on-dot electron-electron Coulomb interaction energy. The couplings between the dot and the noninteracting leads are fully captured by the hybridization functions, which assume a Lorentzian form of $\Gamma_\alpha(\omega) = \Delta_\alpha W_\alpha^2/[(\omega - \Omega_\alpha)^2 + W_\alpha^2]$. Here, $\Delta_\alpha$ is the effective dot-lead coupling strength, and $\Omega_\alpha$ and $W_\alpha$ are the band center and width of lead–$\alpha$, respectively.

As for the measured local observable $O$ in Eq. (1), we choose to examine two local operators, the local magnetic susceptibility $\chi^m \equiv \frac{\partial \langle \hat{m}_z \rangle}{\partial H_z}|_{H_z \to 0}$ and the local charge susceptibility $\chi^c \equiv -\frac{\partial \langle \hat{n}_d \rangle}{\partial \epsilon_d}$. Here, $\hat{m}_z = g\mu_{\rm B}(\hat{n}_\uparrow - \hat{n}_\downarrow)/2$ is the dot magnetization operator, $H_z$ the magnetic field, $g$ the gyromagnetic ratio, and $\mu_{\rm B}$ the Bohr magneton.

*The noninteracting case.* For a noninteracting dot ($U = 0$) coupled to wide-band leads ($W_\alpha \to \infty$), $\chi^m$ and $\chi^c$ can be expressed in compact form as [45]

$$O = \mathcal{C}_O \sum_\alpha \Delta_\alpha \int d\omega\, \frac{\partial A(\omega)}{\partial \epsilon_d}\, f_{T_\alpha,\mu_\alpha}(\omega). \tag{2}$$

Here, $f_{T_\alpha,\mu_\alpha}(\omega) = 1/[e^{(\omega-\mu_\alpha)/T_\alpha} + 1]$ is the Fermi function, $A(\omega)$ is the dot spectral function [45], and $\mathcal{C}_O$ is a constant prefactor dependent on the specific choice of $O$. The perturbation of local observable $O$ by the coupled probe assumes the following general form:

$$\delta O_p(T_p, \mu_p) = -\mathcal{C}_O\, \Delta_p \int d\omega\, \frac{\partial A(\omega)}{\partial \epsilon_d} \Big\{ f_{T_p,\mu_p}(\omega) - \frac{\Delta_L f_{T_L,\mu_L}(\omega) + \Delta_R f_{T_R,\mu_R}(\omega)}{\Delta_L + \Delta_R} \Big\}. \tag{3}$$

It is then obvious that, in this noninteracting case, applying the MPC to $\chi^m$ or $\chi^c$ would lead to exactly the same $T^*$.

Let us now examine in detail how the excitations induced by a bias voltage or thermal gradient affect the local observable $O$. Denote with $O_0(T_L, \mu_L, T_R, \mu_R)$ the expectation value of $\hat{O}$ in the *absence* of the probe, and its deviation from the equilibrium value is

$$\begin{aligned} O_0(T_L, \mu_L, T_R, \mu_R) - O_0(T_{\rm eq}, \mu_{\rm eq}, T_{\rm eq}, \mu_{\rm eq}) = \\ -\mathcal{C}_O\,(\Delta_L + \Delta_R) \int d\omega\, \frac{\partial A(\omega)}{\partial \epsilon_d} \Big\{ f_{T_{\rm eq},\mu_{\rm eq}}(\omega) \\ - \frac{\Delta_L f_{T_L,\mu_L}(\omega) + \Delta_R f_{T_R,\mu_R}(\omega)}{\Delta_L + \Delta_R} \Big\}. \end{aligned} \tag{4}$$

Here, $T_{\rm eq}$ and $\mu_{\rm eq}$ are the temperature and chemical potential of the QD at equilibrium, respectively.

By comparing Eqs. (3) and (4), we immediately recognize that

$$O_0(T_L, \mu_L, T_R, \mu_R) = O_0(T^*, \mu^*, T^*, \mu^*), \tag{5}$$

provided

$$\left.\frac{\delta O_p(T_p, \mu_p)}{\Delta_p}\right|_{T_p = T^*,\, \mu_p = \mu^*,\, \Delta_p \to 0} = 0 \tag{6}$$

can be achieved. In relation to Eq. (1), Eq. (6) further requires that the perturbation to the local observable $O$ by the coupled probe minimizes to zero.

Equation (5) is the central result of this work. As illustrated in Fig. 1, it establishes a quantitative relation between the local property of a nonequilibrium dot and that of a reference equilibrium dot. The physical significance of $T^*$ is thus clarified: the electronic excitations induced by a bias voltage or temperature gradient can be equivalently characterized as thermal excitations induced by a uniform equilibrium temperature. This then provides a *microscopic* interpretation of the MPC-based definition of local temperature.

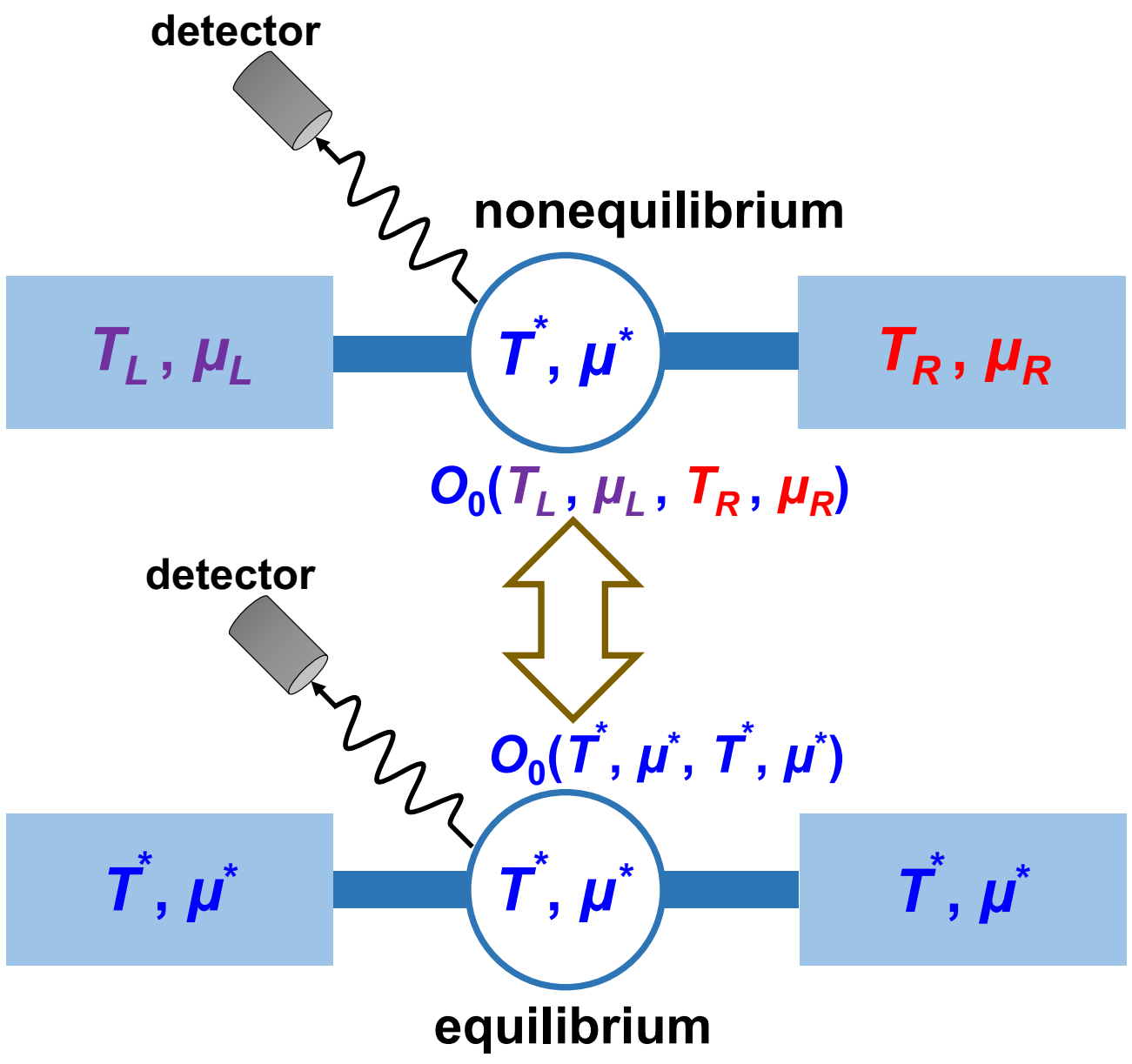

FIG. 1. Schematic illustration of Eq. (5). The local observable $O_0$ of a nonequilibrium QD can be made equivalent to that of a reference equilibrium QD, provided the two dots have the same local temperature $T^*$.

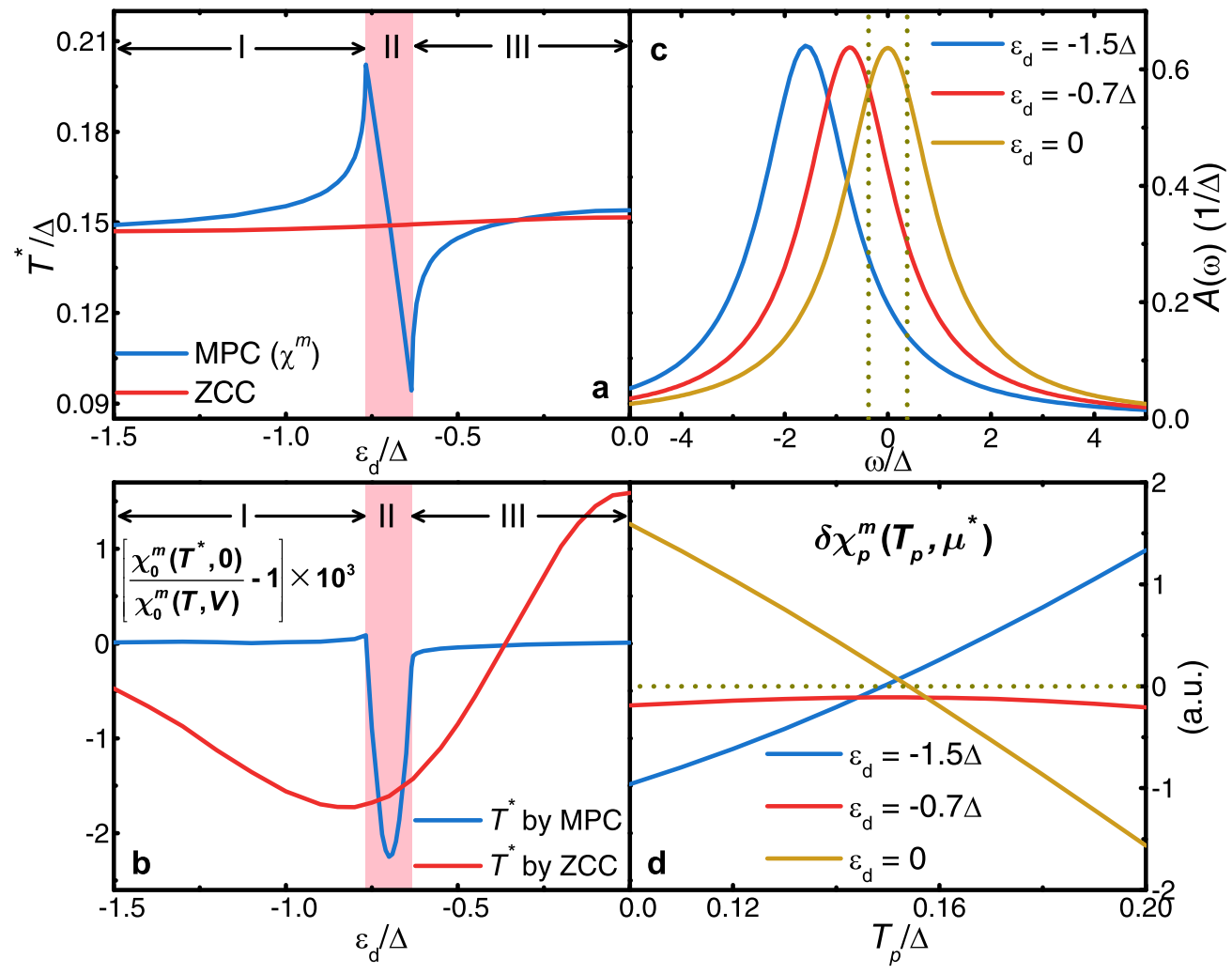

FIG. 2. Calculated (a) $T^*$ and (b) relative deviation between $\chi_0^m(T,V)$ and $\chi_0^m(T^*,0)$ versus $\epsilon_d$ for a noninteracting QD under a bias voltage $V$. (c) Dot spectral function $A(\omega)$ and (d) $\delta\chi_p^m$ versus $T_p$ for different $\epsilon_d$. The QD parameters are (in units of $\Delta$): $U = 0$, $T_L = T_R = T = 0.1$, $\mu_R = -\mu_L = V/2 = 0.2$, $\Delta_L = \Delta_R = 0.5$, $\Omega_L = \Omega_R = 0$, and $W_L = W_R = 20$. The vertical lines and regions are explained in the main text.

In contrast, within the same conditions, the ZCC has a distinct different form from that in Eq. (4) [45, 47], suggesting that in general the local temperature $T^*$ determined by the ZCC does not guarantee the equality in Eq. (5).

To verify the above analytical analysis, and to demonstrate that Eq. (5) underscores the physical significance of $T^*$, we perform numerical calculations on the Anderson model with an accurate and universal hierarchical equations of motion (HEOM) approach [48–50]. The HEOM theory is in principle formally exact, and in practice the numerical results converge rapidly and uniformly with the truncation tier of the hierarchy [45].

Figure 2(a) shows $T^*$ determined by the MPC and the ZCC for a noninteracting QD of varying $\epsilon_d$ under a fixed bias voltage. The ZCC predicts an almost constant $T^*$ over a large range of $\epsilon_d$. In contrast, the MPC results in a conspicuous fluctuation of $T^*$ around $\epsilon_d = -0.7\Delta$ ($\Delta = \Delta_L + \Delta_R$ is taken as the unit of energy), where the magnitude of $T^*$ deviates significantly from the ZCC value. The vertical lines in Fig. 2(a) enclose a region (region II) in which the MPC-determined $T^*$ varies sharply with increasing $\epsilon_d$. In this region the zero-perturbation condition, Eq. (6), is out of reach no matter how $T_p$ is varied, while in the other regions (I and III) Eq. (6) is satisfied for any $\epsilon_d$, as exemplified in Fig. 2(d).

We now examine the correspondence relation for local properties, which is cast in a simplified form of $O_0(T,V) = O_0(T^*,0)$ in the case $T_L = T_R = T$. Figure 2(b) shows the relative deviation $[\chi_0^m(T^*,0) - \chi_0^m(T,V)]/\chi_0^m(T,V)$ obtained numerically. With $T^*$ determined by the MPC, such deviation appears to be vanishingly small in regions I and III, where the zero perturbation condition, Eq. (6), is always achievable; while in region II the deviation remains appreciable. This clearly verifies our analytical conclusion that the zero perturbation condition for determining $T^*$ is a prerequisite for the correspondence relation to hold. In contrast, with $T^*$ determined by the ZCC, the correspondence relation does not apply over the large range of $\epsilon_d$ examined.

The existence of the three distinct regions for the MPC-determined $T^*$ can be understood as follows. As shown in Fig. 2(c), the dot spectral function $A(\omega)$ exhibits a single peak centered at $\epsilon_d$ and broadened by $\Delta$. Under a finite voltage, most of the electronic excitations occur within an energy window $(\mu_L-\omega_L, \mu_R+\omega_R)$, where $\omega_\alpha$ is the full width at half maximum of $\frac{\partial}{\partial\omega} f_{T_\alpha,\mu_\alpha}(\omega)$. For a dot in region I (such as $\epsilon_d = -1.5\Delta$), the bulk of the dot spectral peak lies below the excitation window, and the dot level is off-resonant with the lead states. Consequently, the electronic excitations are largely local on the dot, and the MPC-determined $T^*$ precisely captures the magnitudes of these local excitations.

In contrast, for a dot in region III (such as $\epsilon_d = 0$), the center of the dot spectral peak resides precisely in the excitation window, indicating that the dot level is in strong resonance with the lead states. Therefore, excitations occur mostly inside the leads to create hot electrons (holes) above (below) the Fermi energy, and hence are rather *nonlocal*. In such a case, the MPC-determined $T^*$ quantifies the magnitude of these nonlocal excitations. In this respect, it is more appropriate to interpret $T^*$ as an "effective temperature", rather than a local temperature.

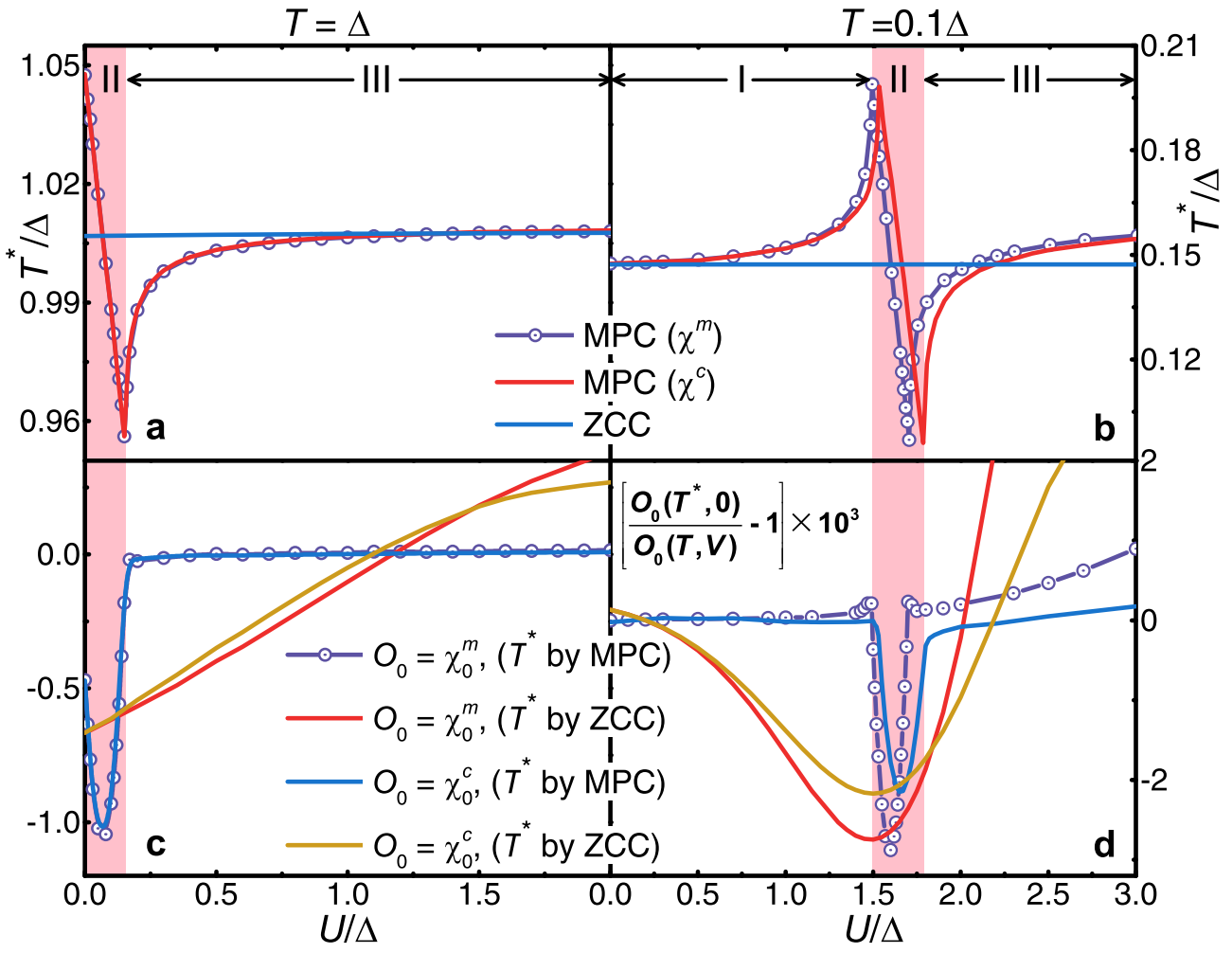

FIG. 3. Calculated $T^*$ versus $U$ for an interacting QD under a bias voltage $V$ at the background temperature (a) $T = \Delta$ and (b) $T = 0.1\Delta$ [45]. The relative deviations between $O_0(T, V)$ and $O_0(T^*, 0)$ are shown in (c) and (d), respectively. The QD parameters are (in units of $\Delta$): $\epsilon_d = -2$, $\mu_R = -\mu_L = V/2 = 0.2$, $\Delta_L = \Delta_R = 0.5$, $\Omega_L = \Omega_R = 0$, and $W_L = W_R = 20$. The vertical lines and regions are explained in the main text.

Finally, for a dot in region II (such as $\epsilon_d = -0.7\Delta$), the spectral peak lies at the edge of the excitation window. The dot is thus in a near-resonance situation, and local and nonlocal excitations could both take place. Since the local and nonlocal excitations are intrinsically different, their influence on the local properties cannot be adequately addressed by a single thermodynamic parameter $T^*$. This thus explains why the zero-perturbation condition of Eq. (6) is out of reach in region II.

*The interacting case.* Analytical analysis is somewhat difficult for interacting QDs, and therefore we resort to a numerical analysis by employing the HEOM approach. For a QD with $U > 0$, $\chi^m = -\frac{1}{2}g^2\mu_B^2\left(\frac{\partial\langle n_\uparrow\rangle}{\partial\epsilon_\uparrow} - \frac{\partial\langle n_\downarrow\rangle}{\partial\epsilon_\uparrow}\right)$ and $\chi^c = -2\left(\frac{\partial\langle n_\uparrow\rangle}{\partial\epsilon_\uparrow} + \frac{\partial\langle n_\downarrow\rangle}{\partial\epsilon_\uparrow}\right)$ are nonequivalent because $\frac{\partial\langle n_\downarrow\rangle}{\partial\epsilon_\uparrow} \neq 0$. Therefore, the MPC-determined local temperature depends on the specific choice of local observable $O$. Nevertheless, our calculations have shown that over a wide range of parameters the use of $\chi^m$ and $\chi^c$ result in similar values of $T^*$ (see Fig. 3).

Figure 3(a) and (b) depict the variation of $T^*$ with increasing $U$ at a high ($T = \Delta$) and low ($T = 0.1\Delta$) background temperature, respectively. Similar to the noninteracting situation, the ZCC predicts an almost constant $T^*$ over a large range of $U$, while the MPC again gives rise to a sharp transition of $T^*$ within a small region (region II) of $U$. The relative deviations $[O_0(T^*, 0) - O_0(T, V)]/O_0(T, V)$ ($O = \chi^m$ and $\chi^c$) are shown in Fig. 3(c) and (d). While the ZCC-defined $T^*$ does not conform to the correspondence relation of Eq. (5), the MPC-determined $T^*$ leads to rather minor deviations so long as the zero perturbation of Eq. (6) can be achieved (in regions I and III). Here, the regions I, II, and III correspond to the off-resonant, near-resonant, and resonant situations, respectively, as supported by the positions of the dot spectral peaks with respect to the excitation energy window [45].

At a low background temperature (such as $T = 0.1\Delta$) the deviation between $\chi_0^m(T, V)$ and $\chi_0^m(T^*, 0)$ (with the MPC-determined $T^*$) assumes a small but finite value in region III; see Fig. 3(d). This is because Kondo resonant states start to form as $U$ increases. Under a bias voltage, Kondo resonant states facilitate electron co-tunneling processes, which can be understood as the concurrence of local spin-flip and nonlocal electron-transfer excitations. As in the case of noninteracting electrons, such mixed-ranged excitations cannot be fully captured by the single parameter $T^*$, and hence the correspondence relation for local observables does not hold.

*Conclusions.* Based on the above analysis, we can now answer the question of how to physically interpret the defined or measured local temperature. The ZCC-based definition does give a $T^*$ that is higher than the background $T$, indicating the presence of local heating. However, the magnitude of $T^*$ can hardly be associated with the changes in system local observables. In contrast, the MPC-based definition establishes a quantitative correspondence between the nonequilibrium system of interest and a reference equilibrium system. The correspondence relation holds as long as the following three conditions are met: (i) the perturbation induced by the probe minimizes to zero; (ii) the monitored observable is a local property; and (iii) the electronic excitations driven by the external source are fully local.

Finally, from the experimental perspective the MPC-based definition is obviously more practical, as it does not require measuring the heat currents. Even if the zero perturbation to a local observable is out of reach, the MPC of Eq. (1) always provides a definitive measurement for the magnitude of $T^*$. The nonzero minimal perturbation indicates the presence of a nonlocal contribution to local heating from electronic excitations in the environment. In view of the fact that local thermal probes as those suggested in this work are now being developed, we hope our studies will provide a firmer basis for understanding the ensuing quantities they measure.

The support from the National Natural Science Foundation of China (Grants No. 21233007, No. 21322305, and No. 21573202), the Fundamental Research Funds for the Central Universities (Grants No. 2030020028 and No. 2340000074), the Ministry of Science and Technology (Grant No. 2016YFA0400900), and the Strategic Priority Research Program (B) of the Chinese Academy of Sciences (Grant No. XDB01020000) is gratefully ap-preciated. M. D. acknowledges support from Department of Energy under Grant No. DE-FG02-05ER46204.

* xz58@ustc.edu.cn
† diventra@physics.ucsd.edu

**Supplementary Information for**

# The thermodynamic meaning of local temperature of nonequilibrium open quantum systems

LvZhou Ye,[1] Xiao Zheng,[1,2,*] YiJing Yan,[1,3] and Massimiliano Di Ventra[4,†]

[1]*Hefei National Laboratory for Physical Sciences at the Microscale, University of Science and Technology of China, Hefei, Anhui 230026, China*

[2]*Synergetic Innovation Center of Quantum Information and Quantum Physics, CAS Center for Excellence in Nanoscience, University of Science and Technology of China, Hefei, Anhui 230026, China*

[3]*iChEM (Collaborative Innovation Center of Chemistry for Energy Materials), University of Science and Technology of China, Hefei, Anhui 230026, China*

[4]*Department of Physics, University of California, San Diego, La Jolla, California 92093*

## CONTENTS

## I. ZERO-CURRENT CONDITION

Within the zero-current condition (ZCC), an ideal potentiometer/thermometer (the probe) is coupled to the nonequilibrium system of interest. By varying the temperature ($T_p$) and chemical potential ($\mu_p$) of the probe until *both* the electric current ($I_p$) and heat current ($J_p^{\mathrm{H}}$) flowing through the probe vanish, the local temperature $T^*$ and local chemical potential $\mu^*$ of the system are determined as $T^* = T_p$ and $\mu^* = \mu_p$, respectively. In short, the ZCC is expressed as (for brevity we set $e = \hbar = k_{\mathrm{B}} = 1$ hereafter)

$$I_p(T_p = T^*, \mu_p = \mu^*) = 0, \tag{S1a}$$

$$J_p^{\mathrm{H}}(T_p = T^*, \mu_p = \mu^*) = 0. \tag{S1b}$$

## II. MINIMAL-PERTURBATION CONDITION

Consider a quantum dot (QD) connected to two leads ($L$ and $R$), with the lead temperatures (chemical potentials) being $T_L$ and $T_R$ ($\mu_L$ and $\mu_R$), respectively. Within the MPC, the local temperature $T^*$ of the QD is determined by varying $T_p$ so that the perturbation $\delta O_p(T_p, \mu_p)$ due to the lead-dot coupling is minimized [1, 2]:

$$T^* = \arg\min_{T_p} |\delta O_p(T_p, \mu_p = \mu^*)| \,. \tag{S2}$$

Here, $\mu_p$ is kept aligned with a preset local chemical potential $\mu^*$ during the variation of $T_p$. For a QD subjected to a bias voltage $V = \mu_R - \mu_L$, the local chemical potential is determined as [2]

$$\mu^* = \zeta_L \mu_L + \zeta_R \mu_R. \tag{S3}$$

The perturbation term in equation (S2) assumes the form of

$$\delta O_p(T_p, \mu_p) = \zeta_L O_p(T_L, \mu_L) + \zeta_R O_p(T_R, \mu_R) - O_p(T_p, \mu_p), \tag{S4}$$

where $O_p(T_\alpha, \mu_\alpha)$ denotes the local observable $\langle \hat{O} \rangle$ measured by setting $T_p = T_\alpha$ and $\mu_p = \mu_\alpha$ ($\alpha = L$ or $R$); and the weight coefficients $\zeta_\alpha$ ($\alpha = L, R$) are determined by

$$\zeta_\alpha = 1 - \left| \frac{I_p(T_\alpha, \mu_\alpha)}{I_p(T_L, \mu_L) - I_p(T_R, \mu_R)} \right| . \tag{S5}$$

Here, $I_p(T_\alpha, \mu_\alpha)$ is the electric current measured at the probe, by setting the chemical potential and temperature of the probe to be identical with those of lead $\alpha$.

## III. HIERARCHICAL EQUATIONS OF MOTION APPROACH FOR QUANTUM IMPURITY SYSTEMS

We consider a QD described by the single-impurity Anderson model (SIAM) [3]. The total Hamiltonian is $\hat{H} = \hat{H}_{\mathrm{dot}} + \hat{H}_{\mathrm{lead}} + \hat{H}_{\mathrm{coup}}$. The dot is represented by $\hat{H}_{\mathrm{dot}} = \epsilon_d \left(\hat{n}_\uparrow + \hat{n}_\downarrow\right) + U \hat{n}_\uparrow \hat{n}_\downarrow$, as described in the main text. $\hat{H}_{\rm lead} = \sum_{\alpha k s} \epsilon_{\alpha k}\, \hat{d}^\dagger_{\alpha k s}\, \hat{d}_{\alpha k s}$ represents the noninteracting leads, and $\hat{H}_{\rm coup} = \sum_{\alpha k s} (t_{\alpha k}\, \hat{a}^\dagger_s\, \hat{d}_{\alpha k s} + {\rm H.c.})$ describes the dot-lead couplings, respectively. Here, $\hat{d}^\dagger_{\alpha k s}$ ($\hat{d}_{\alpha k s}$) creates (annihilates) a spin–$s$ electron on the state $|k\rangle$ of lead $\alpha$ ($\alpha = L$, $R$ or $p$), and $t_{\alpha k}$ is the coupling strength between the dot level and lead orbital $|k\rangle$. The influence of noninteracting leads can be captured by the hybridization functions $\Gamma_\alpha(\omega) \equiv \pi \sum_k |t_{\alpha k}|^2 \delta(\epsilon - \epsilon_{\alpha k})$. For numerical convenience, a Lorentzian form of $\Gamma_\alpha(\omega) = \Delta_\alpha W_\alpha^2/[(\omega - \Omega_\alpha)^2 + W_\alpha^2]$ is adopted in the main text.

We perform numerical calculations on the SIAM with an accurate and universal hierarchical equations of motion (HEOM) approach [4–6]. The HEOM theory is formally rigorous, and the numerical approach has been routinely used to investigate the equilibrium and nonequilibrium properties of strongly-correlated quantum impurity systems. The derivation and practicality of the HEOM approach have been detailed in Refs. [5] and [6]. Here, we briefly introduce some of its key features. The final form of HEOM can be cast into a compact form as follows [4]

$$\dot{\rho}^{(n)}_{j_1\cdots j_n} = -\Big(i\mathcal{L} + \sum_{r=1}^{n} \gamma_{j_r}\Big)\rho^{(n)}_{j_1\cdots j_n} - i \sum_j \mathcal{A}_{\bar{j}}\, \rho^{(n+1)}_{j_1\cdots j_n j} - i \sum_{r=1}^{n} (-)^{n-r}\, \mathcal{C}_{j_r}\, \rho^{(n-1)}_{j_1\cdots j_{r-1} j_{r+1}\cdots j_n}. \tag{S6}$$

Here, $\rho^{(0)}(t) = \rho(t) \equiv {\rm tr}_{\rm env}\, \rho_{\rm total}(t)$ is the reduced density matrix, and $\{\rho^{(n)}_{j_1\cdots j_n}(t); n = 1, \cdots, N_{\rm trun}\}$ are the auxiliary density matrices, with $N_{\rm trun}$ denoting the truncation level of the hierarchy. The multi-component index $j \equiv (\sigma\alpha\nu\nu' m)$ characterizes the transfer of an electron from/to ($\sigma = +/-$) the impurity level $\nu$ to/from level $\nu'$ via lead $\alpha$, associated with a characteristic memory time $\gamma_m^{-1}$. The Grassmann superoperators $\mathcal{A}_{\bar{j}} \equiv \mathcal{A}^{\bar{\sigma}}_\nu$ and $\mathcal{C}_j \equiv \mathcal{C}^\sigma_{\nu\nu' m}$ are defined via their fermionic/bosonic actions on an operator $\hat{O}$ as $\mathcal{A}^{\bar{\sigma}}_\nu \hat{O} \equiv [\hat{a}^{\bar{\sigma}}_\nu, \hat{O}]_\mp$ and $\mathcal{C}^\sigma_{\nu\nu' m}\hat{O} \equiv \eta^\sigma_{\nu\nu' m}\hat{a}^\sigma_\nu \hat{O} \pm (\eta^{\bar{\sigma}}_{\nu\nu' m})^* \hat{O}\hat{a}^\sigma_{\nu'}$, respectively, with $\bar{\sigma}$ being the opposite sign of $\sigma$. The on-dot electron-electron interactions are included in the Liouvillian of the impurity, $\mathcal{L}\cdot \equiv [\hat{H}_{\rm dot}, \cdot\,]$.

Numerical results of the HEOM method are guaranteed to be quantitatively accurate provided they converge with respect to the truncation level of the hierarchy $N_{\rm trun}$. For the noninteracting (interacting) QDs studied in this work, the convergence is achieved at $N_{\rm trun} = 2$ (4), unless otherwise specified.

## IV. ANALYTICAL ANALYSIS OF LOCAL TEMPERATURE OF A SINGLE-LEVEL QUANTUM DOT

We investigate a single-level QD system in contact with noninteracting leads. By using the nonequilinbrium Green's functions (NEGF) method, the retarded/advanced single-electron Green's function (spin–$s$) $G^{r/a}_s(\omega)$ at stationary state is

$$G^r_s(\omega) = [G^a_s(\omega)]^\dagger = \frac{1}{\omega - \epsilon_d - \Sigma^r_s(\omega)}. \tag{S7}$$

Here, the retarded self-energy $\Sigma^r_s(\omega) = \Sigma^r_{\rm res}(\omega) + \Sigma^r_{\rm ee}(\omega)$, with $\Sigma^r_{\rm res}(\omega)$ being the embedding self-energy due to the lead-dot couplings, and $\Sigma^r_{\rm ee}(\omega)$ being the interacting part due to the electron-electron ($e$-$e$) interactions. Let $\Sigma^r_s(\omega) \equiv B_s(\omega) + iD_s(\omega)$, with $B_s(\omega) = {\rm Re}[\Sigma^r_s(\omega)]$ and $D_s(\omega) =$

$\mathrm{Im}[\Sigma_s^r(\omega)]$. The dot spectral function $A_s(\omega)$ is

$$A_s(\omega) \equiv \frac{1}{2\pi}\int dt\, e^{i\omega t}\langle\{\hat{a}_s(t), \hat{a}_s^\dagger(0)\}\rangle \tag{S8}$$

$$= -\frac{1}{\pi}\mathrm{Im}[G_s^r(\omega)] = -\frac{1}{\pi}\frac{D_s(\omega)}{[\omega-\epsilon_d-B_s(\omega)]^2+[D_s(\omega)]^2}. \tag{S9}$$

The lesser Green's function $G_s^<(\omega)$ is

$$G_s^<(\omega) = G_s^r(\omega)\Sigma_s^<(\omega)G_s^a(\omega) = \frac{\Sigma_s^<(\omega)}{[\omega-\epsilon_d-B_s(\omega)]^2+[D_s(\omega)]^2} = -\pi A_s(\omega)\frac{\Sigma_s^<(\omega)}{D_s(\omega)}, \tag{S10}$$

where the lesser self-energy $\Sigma_s^<(\omega) = \Sigma_{\mathrm{res}}^<(\omega) + \Sigma_{\mathrm{ee}}^<(\omega)$ consists of the embedding $[\Sigma_{\mathrm{res}}^<(\omega)]$ and interacting $[\Sigma_{\mathrm{ee}}^<(\omega)]$ parts.

The occupation number of spin–$s$ electrons on the dot $n_s = \langle\hat{n}_s\rangle$ is evaluated as

$$n_s = \frac{1}{2\pi i}\int d\omega\, G_s^<(\omega) = \frac{i}{2}\int d\omega\, A_s(\omega)\frac{\Sigma_s^<(\omega)}{D_s(\omega)}. \tag{S11}$$

The energy distribution of electric and heat currents flowing into lead $\alpha$ is [7]

$$j_{\alpha s}^k(\omega) = (-1)^{k+1}\frac{i}{\pi}(\omega-\mu_\alpha)^k\,\Gamma_\alpha(\omega)\left\{G_s^<(\omega) + 2if_{T_\alpha,\mu_\alpha}(\omega)\mathrm{Im}[G_s^r(\omega)]\right\}. \tag{S12}$$

Here, $k = 0$ and 1 correspond to the electric and heat currents, respectively. The electric and heat currents are then calculated as $I_\alpha = \sum_s\int d\omega\, j_{\alpha s}^0(\omega)$ and $J_\alpha^{\mathrm{H}} = \sum_s\int d\omega\, j_{\alpha s}^1(\omega)$, respectively.

### A. The noninteracting dot

For noninteracting dots ($U = 0$), the interacting parts of the self-energies $[\Sigma_{\mathrm{ee}}^r(\omega)$ and $\Sigma_{\mathrm{ee}}^<(\omega)]$ vanish. In the main text, we consider all the leads have the same bandwidth $W_\alpha = W$ ($\alpha = L, R$ and $p$). The hybridization function is

$$\Gamma_\alpha(\omega) = \frac{\Delta_\alpha W^2}{(\omega-\Omega_\alpha)^2+W^2} = \Delta_\alpha\eta_\alpha(\omega). \tag{S13}$$

Here, the function $\eta_\alpha(\omega) \equiv W^2/\left[(\omega-\Omega_\alpha)^2+W^2\right]$ is proportional to the density of states of lead $\alpha$. The lead band center is set to the chemical potential $\Omega_\alpha = \mu_\alpha$. We then have

$$\Sigma_s^r(\omega) = \sum_\alpha\frac{\Delta_\alpha W}{\omega-\mu_\alpha+iW} = B_s(\omega) + iD_s(\omega), \tag{S14}$$

$$\Sigma_s^<(\omega) = 2i\sum_\alpha\Gamma_\alpha(\omega)f_{T_\alpha,\mu_\alpha}(\omega), \tag{S15}$$

where

$$B_s(\omega) = \sum_\alpha\Gamma_\alpha(\omega)\frac{\omega-\mu_\alpha}{W}, \tag{S16}$$

$$D_s(\omega) = -\sum_\alpha\Gamma_\alpha(\omega). \tag{S17}$$

The dot spectral function and lesser Green's function are

$$A_s(\omega) = \frac{1}{\pi}\frac{\sum_\alpha \Gamma_\alpha(\omega)}{[\omega - \epsilon_d - B_s(\omega)]^2 + [D_s(\omega)]^2}, \tag{S18}$$

$$G_s^<(\omega) = 2\pi i A_s(\omega)\frac{\sum_\alpha \Gamma_\alpha(\omega) f_{T_\alpha,\mu_\alpha}(\omega)}{\sum_\alpha \Gamma_\alpha(\omega)}. \tag{S19}$$

The electric current is

$$I_{\alpha s} = 2\int d\omega\, \Gamma_\alpha(\omega) A_s(\omega)\left\{\frac{\sum_{\alpha'} \Gamma_{\alpha'}(\omega) f_{T_{\alpha'},\mu_{\alpha'}}(\omega)}{\sum_{\alpha'} \Gamma_{\alpha'}(\omega)} - f_{T_\alpha,\mu_\alpha}(\omega)\right\}. \tag{S20}$$

To determine the weight coefficients $\{\zeta_\alpha\}$, we calculate the electric currents through the probe by setting $T_p = T_\alpha$ and $\mu_p = \mu_\alpha$ ($\alpha = L$ or $R$):

$$I_p(T_L,\mu_L) = -2\int d\omega\, \frac{\Delta_R \Delta_p\, \eta_L(\omega)\, \eta_R(\omega)}{(\Delta_L + \Delta_p)\eta_L(\omega) + \Delta_R \eta_R(\omega)}\left[f_{T_L,\mu_L}(\omega) - f_{T_R,\mu_R}(\omega)\right] A(\omega)\big|_{T_p=T_L,\, \mu_p=\mu_L}, \tag{S21}$$

$$I_p(T_R,\mu_R) = -2\int d\omega\, \frac{\Delta_L \Delta_p\, \eta_L(\omega)\, \eta_R(\omega)}{\Delta_L \eta_L(\omega) + (\Delta_R + \Delta_p)\eta_R(\omega)}\left[f_{T_R,\mu_R}(\omega) - f_{T_L,\mu_L}(\omega)\right] A(\omega)\big|_{T_p=T_R,\, \mu_p=\mu_R}. \tag{S22}$$

Here, $A(\omega) = \sum_s A_s(\omega)$ is the total dot spectral function. At $\Delta_p = 0$, we have $A(\omega) = A_0(\omega)$, with $A_0(\omega)$ being the dot spectral function in the absence of the probe. It is straightforward to see that

$$\left.\frac{I_p(T_L,\mu_L)}{I_p(T_R,\mu_R)}\right|_{\Delta_p\to 0} = -\frac{\Delta_R}{\Delta_L}. \tag{S23}$$

The weight coefficients are then determined as

$$\zeta_L = \frac{\Delta_L}{\Delta_L + \Delta_R} \quad \text{and} \quad \zeta_R = \frac{\Delta_R}{\Delta_L + \Delta_R}. \tag{S24}$$

Therefore, the local chemical potential is

$$\mu^* = \frac{\Delta_L}{\Delta_L + \Delta_R}\mu_L + \frac{\Delta_R}{\Delta_L + \Delta_R}\mu_R. \tag{S25}$$

We then evaluate the local observables. The electron occupation number is [cf. equation (S11)]

$$n_s = \sum_\alpha \int d\omega\, \frac{\Gamma_\alpha(\omega)}{\sum_{\alpha'} \Gamma_{\alpha'}(\omega)} A_s(\omega) f_{T_\alpha,\mu_\alpha}(\omega). \tag{S26}$$

The local observables $O = \chi^c$ and $\chi^m$ can thus be expressed as

$$O = \mathcal{C}'_O \sum_\alpha \int d\omega\, \frac{\Gamma_\alpha(\omega)}{\sum_{\alpha'} \Gamma_{\alpha'}(\omega)} \frac{\partial A(\omega)}{\partial \epsilon_d} f_{T_\alpha,\mu_\alpha}(\omega). \tag{S27}$$

Here, $\mathcal{C}'_O$ is a constant prefactor dependent on the specific choice of $O$. By setting $T_p = T_\alpha$ and $\mu_p = \mu_\alpha$ ($\alpha = L$ or $R$), we have

$$O_p(T_L,\mu_L) = \mathcal{C}'_O \int d\omega\, \frac{(\Delta_L + \Delta_p)\eta_L(\omega) f_{T_L,\mu_L}(\omega) + \Delta_R\, \eta_R(\omega) f_{T_R,\mu_R}(\omega)}{(\Delta_L + \Delta_p)\eta_L(\omega) + \Delta_R\, \eta_R(\omega)} \left.\frac{\partial A(\omega)}{\partial \epsilon_d}\right|_{T_p=T_L,\, \mu_p=\mu_L}, \tag{S28}$$

$$O_p(T_R,\mu_R) = \mathcal{C}'_O \int d\omega\, \frac{\Delta_L\, \eta_L(\omega) f_{T_L,\mu_L}(\omega) + (\Delta_R + \Delta_p)\eta_R(\omega) f_{T_R,\mu_R}(\omega)}{\Delta_L\, \eta_L(\omega) + (\Delta_R + \Delta_p)\eta_R(\omega)} \left.\frac{\partial A(\omega)}{\partial \epsilon_d}\right|_{T_p=T_R,\, \mu_p=\mu_R}. \tag{S29}$$

Let us compare a noninteracting QD in the nonequilibrium state characterized by the parameters $(T_L, \mu_L, T_R, \mu_R)$ and that in an equilibrium state of $(T_{\rm eq}, \mu_{\rm eq}, T_{\rm eq}, \mu_{\rm eq})$:

$$\begin{aligned} &O_0(T_L, \mu_L, T_R, \mu_R) - O_0(T_{\rm eq}, \mu_{\rm eq}, T_{\rm eq}, \mu_{\rm eq}) \\ &= \mathcal{C}'_O \int d\omega \left\{ \frac{\Delta_L\, \eta_L(\omega) f_{T_L,\mu_L}(\omega) + \Delta_R\, \eta_R(\omega) f_{T_R,\mu_R}(\omega)}{\Delta_L\, \eta_L(\omega) + \Delta_R\, \eta_R(\omega)} \frac{\partial A_0(\omega)}{\partial \epsilon_d} \right. \\ &\left. - f_{T_{\rm eq},\mu_{\rm eq}}(\omega) \left. \frac{\partial A_0(\omega)}{\partial \epsilon_d} \right|_{T_\alpha = T_{\rm eq},\, \mu_\alpha = \mu_{\rm eq}} \right\}. \end{aligned} \tag{S30}$$

Until now, all the derivations above are formally exact for any noninteracting QD system.

### *1. The wide-band limit case*

In the wide-band limit ($W \to \infty$), $\eta_\alpha(\omega) = 1$, $B_s(\omega) = 0$ and $D_s(\omega) = -\sum_\alpha \Delta_\alpha$. The dot spectral function

$$A_s(\omega) = \frac{1}{\pi} \frac{\sum_\alpha \Delta_\alpha}{(\omega - \epsilon_d)^2 + (\sum_\alpha \Delta_\alpha)^2} \tag{S31}$$

is independent of $\mu_\alpha$ and $T_\alpha$. Consequently, the equilibrium and nonequilibrium dots have identical spectral functions, *i.e.*, $A_0(\omega; \mu_L, \mu_R) = A_0(\omega; \mu_{\rm eq}, \mu_{\rm eq}) = A_0(\omega)$. The perturbation of $O$ by the coupled probe, equation (S4), is

$$\left. \frac{\delta O_p(T_p, \mu_p)}{\Delta_p} \right|_{\Delta_p \to 0} = -\mathcal{C}_O \int d\omega\, \frac{\partial A_0(\omega)}{\partial \epsilon_d} \left\{ f_{T_p,\mu_p}(\omega) - \frac{\Delta_L f_{T_L,\mu_L}(\omega) + \Delta_R f_{T_R,\mu_R}(\omega)}{\Delta_L + \Delta_R} \right\}. \tag{S32}$$

with $\mathcal{C}_O \equiv \frac{\mathcal{C}'_O}{\Delta_L + \Delta_R}$. Equation (S30) becomes

$$\begin{aligned} &O_0(T_L, \mu_L, T_R, \mu_R) - O_0(T_{\rm eq}, \mu_{\rm eq}, T_{\rm eq}, \mu_{\rm eq}) \\ &= -\, \mathcal{C}_O(\Delta_L + \Delta_R) \int d\omega\, \frac{\partial A_0(\omega)}{\partial \epsilon_d} \left\{ f_{T_{\rm eq},\mu_{\rm eq}}(\omega) - \frac{\Delta_L f_{T_L,\mu_L}(\omega) + \Delta_R f_{T_R,\mu_R}(\omega)}{\Delta_L + \Delta_R} \right\}. \end{aligned} \tag{S33}$$

Clearly, if the right-hand side of equation (S32) minimizes to zero at $T_p = T^*$ and $\mu_p = \mu^*$, we immediately have the correspondence relation of

$$O_0(T_L, \mu_L, T_R, \mu_R) = O_0(T^*, \mu^*, T^*, \mu^*). \tag{S34}$$

In contrast, within the same conditions, the ZCC of equation (S1) amounts to [cf. equation (S12)]

$$\int d\omega\, (\omega - \mu^*)^k\, A(\omega) \left\{ f_{T^*,\mu^*}(\omega) - \frac{\Delta_L f_{T_L,\mu_L}(\omega) + \Delta_R f_{T_R,\mu_R}(\omega)}{\Delta_L + \Delta_R} \right\} = 0, \tag{S35}$$

where $k = 0$ and 1 correspond to the electric and heat currents through the probe, respectively. The integral in equation (S35) has a distinct different form than that in equation (S33), suggesting that in general the local temperature $T^*$ determined by the ZCC does not guarantee the equality in equation (S34).

### 2. The finite-band-width case

For leads with a finite band width, the corresponding dot spectral function $A_s(\omega)$ depends explicitly on the lead chemical potential $\mu_\alpha$. Consequently, the nonequilibrium dot spectral function $A_0(\omega;\mu_L,\mu_R)$ differs from the equilibrium counterpart $A_0(\omega;\mu_{\rm eq},\mu_{\rm eq})$. Nevertheless, as will be shown below, the correspondence relation of equation (S34) still holds under a small applied bias voltage $V=\mu_R-\mu_L$.

From equations (S24) and (S25), we have

$$\mu_L = \mu^* - \zeta_R V, \tag{S36}$$

$$\mu_R = \mu^* + \zeta_L V. \tag{S37}$$

The Talyor expasion of $\eta_\alpha(\omega)$ leads to

$$\eta_L(\omega) = \eta(\omega;\mu^*) - 2\zeta_R\left[\frac{\eta(\omega;\mu^*)}{W}\right]^2(\omega-\mu^*)V + O(V^2), \tag{S38}$$

$$\eta_R(\omega) = \eta(\omega;\mu^*) + 2\zeta_L\left[\frac{\eta(\omega;\mu^*)}{W}\right]^2(\omega-\mu^*)V + O(V^2). \tag{S39}$$

Here, $\eta(\omega;\mu^*)\equiv W^2/\left[(\omega-\mu^*)^2+W^2\right]$. We thus have

$$\Delta_L\,\eta_L(\omega) + \Delta_R\,\eta_R(\omega) = (\Delta_L+\Delta_R)\,\eta(\omega;\mu^*) + O(V^2). \tag{S40}$$

We now evaluate $A_0(\omega)$ of equation (S18). The real and imaginary parts of the retarded self-energy, $B_0(\omega)$ and $D_0(\omega)$, are

$$\begin{aligned} B_0(\omega;\mu_L,\mu_R) &= \Delta_L\eta_L(\omega)\frac{\omega-\mu_L}{W} + \Delta_R\eta_R(\omega)\frac{\omega-\mu_R}{W} \\ &= \frac{\omega-\mu^*}{W}(\Delta_L+\Delta_R)\,\eta(\omega;\mu^*) + O(V^2) \\ &= B_0(\omega;\mu^*,\mu^*) + O(V^2), \end{aligned} \tag{S41}$$

$$\begin{aligned} D_0(\omega;\mu_L,\mu_R) &= -\left[\Delta_L\,\eta_L(\omega) + \Delta_R\,\eta_R(\omega)\right] \\ &= -(\Delta_L+\Delta_R)\,\eta(\omega;\mu^*) + O(V^2) \\ &= D_0(\omega;\mu^*,\mu^*) + O(V^2). \end{aligned} \tag{S42}$$

Therefore, we have

$$A_0(\omega;\mu_L,\mu_R) = A_0(\omega;\mu^*,\mu^*) + O(V^2). \tag{S43}$$

The perturbation term equation (S4) at $T_p=T^*$ and $\mu_p=\mu^*$ can be expressed as

$$\begin{aligned} &\left.\frac{\delta O_p(T_p,\mu_p)}{\Delta_p}\right|_{T_p=T^*,\,\mu_p=\mu^*,\,\Delta_p\to 0} \\ &= \mathcal{C}_O'\int d\omega\,\frac{\zeta_L\,\eta_L(\omega)f_{T_L,\mu_L}(\omega)+\zeta_R\,\eta_R(\omega)f_{T_R,\mu_R}(\omega)-\eta(\omega;\mu^*)f_{T^*,\mu^*}(\omega)}{\Delta_L\,\eta_L(\omega)+\Delta_R\,\eta_R(\omega)}\,\frac{\partial A_0(\omega)}{\partial\epsilon_d} + O(V^2). \end{aligned} \tag{S44}$$

Substituting equations (S40) and (S43) into equation (S30) leads to

$$O_0(T_L,\mu_L,T_R,\mu_R)-O_0(T^*,\mu^*,T^*,\mu^*)$$
$$=\mathcal{C}'_O(\Delta_L+\Delta_R)\int d\omega\,\frac{\zeta_L\,\eta_L(\omega)f_{T_L,\mu_L}(\omega)+\zeta_R\,\eta_R(\omega)f_{T_R,\mu_R}(\omega)-\eta(\omega;\mu^*)f_{T^*,\mu^*}(\omega)}{\Delta_L\,\eta_L(\omega)+\Delta_R\,\eta_R(\omega)}\frac{\partial A_0(\omega)}{\partial\epsilon_d}$$
$$+O(V^2). \tag{S45}$$

By comparing equations (S44) and (S45), we have

$$O_0(T_L,\mu_L,T_R,\mu_R)=O_0(T^*,\mu^*,T^*,\mu^*)+O(V^2), \tag{S46}$$

provided that the zero perturbation of equation (S44) can be achieved.

### B. The interacting dot

For QDs with a finite $U$, the interacting parts of the self-energies, $\Sigma^r_{\rm ee}(\omega)$ and $\Sigma^<_{\rm ee}(\omega)$, are dependent on parameters such as Coulomb energy $U$, dot level $\epsilon_s$, temperature $T_\alpha$, and chemical potential $\mu_\alpha$, and thus assume a complicated form. This makes an analytical analysis rather difficult. Therefore, the correspondence relation for interacting QDs are verified numerically by employing the accurate HEOM approach, as described in the main text.

## V. ZCC-DETERMINED $T^*$ AND $\mu^*$

For the QDs studied in the main text, the local chemical potential $\mu^*$ determined by the ZCC is smaller than $10^{-3}\Delta$, while $\mu^*$ determined by the MPC are nearly zero. Therefore, the difference between the ZCC-determined $\mu^*$ and the MPC counterpart is much smaller than the magnitude of applied bias voltage, and is thus considered as a minor difference.

For interacting QDs at a low temperature (such as $T=0.1\Delta$), a somewhat high truncation tier $N_{\rm trun}$ is required to obtain the fully converged heat current $J_p^{\rm H}$ with the HEOM approach. The computational cost is way beyond the computer resources at our disposal. To circumvent this problem, the ZCC-determined $T^*$ and $\mu^*$ of QDs with a finite $U$ assume the values of a noninteracting dot with $U=0$. For the latter, both $I_p$ and $J_p^{\rm H}$ can be obtained very accurately with a low truncation tier. This is also concluded from our extensive numerical simulations that the ZCC-determined $T^*$ and $\mu^*$ tend to vary rather smoothly with the increase of $U$.

As depicted in Fig. S1, the estimated $T^*$ indeed satisfies the zero-electric-current condition. By setting $T_p$ to the (estimated) ZCC-determined $T^*$, the scaled electric current measured at the probe $I_p/\Delta_p$ is vanishingly small as compared to $I_R/\Delta_R$. Moreover, the scaled probe current is also significantly smaller than that measured by setting $T_p$ to the MPC-determined $T^*$, particularly in the near-resonant region. This clearly indicates that the above estimate for the ZCC-determined $T^*$ and $\mu^*$ are rather reasonable.

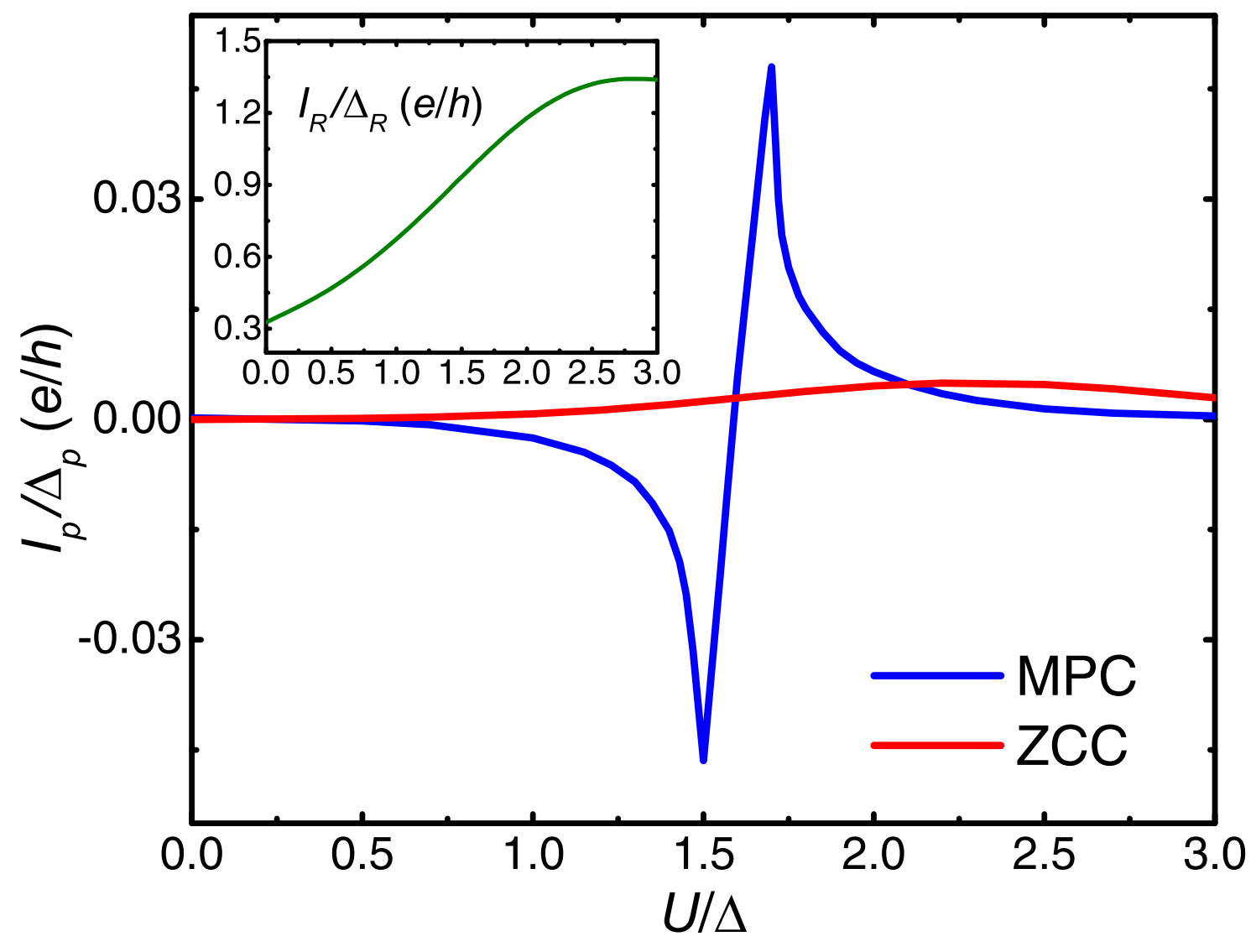

Figure S1. The scaled electric current flowing through the probe $I_p/\Delta_p$ for the QD studied in Fig. 3(b) of the main text when the MPC or ZCC is achieved. The ZCC-determined $T^*$ and $\mu^*$ for QDs with a finite $U$ assume the values of the dot at $U = 0$. The dot-probe coupling strength is $\Delta_p = 0.005\Delta$. The inset shows the scaled electric current flowing through the right lead $I_R/\Delta_R$.

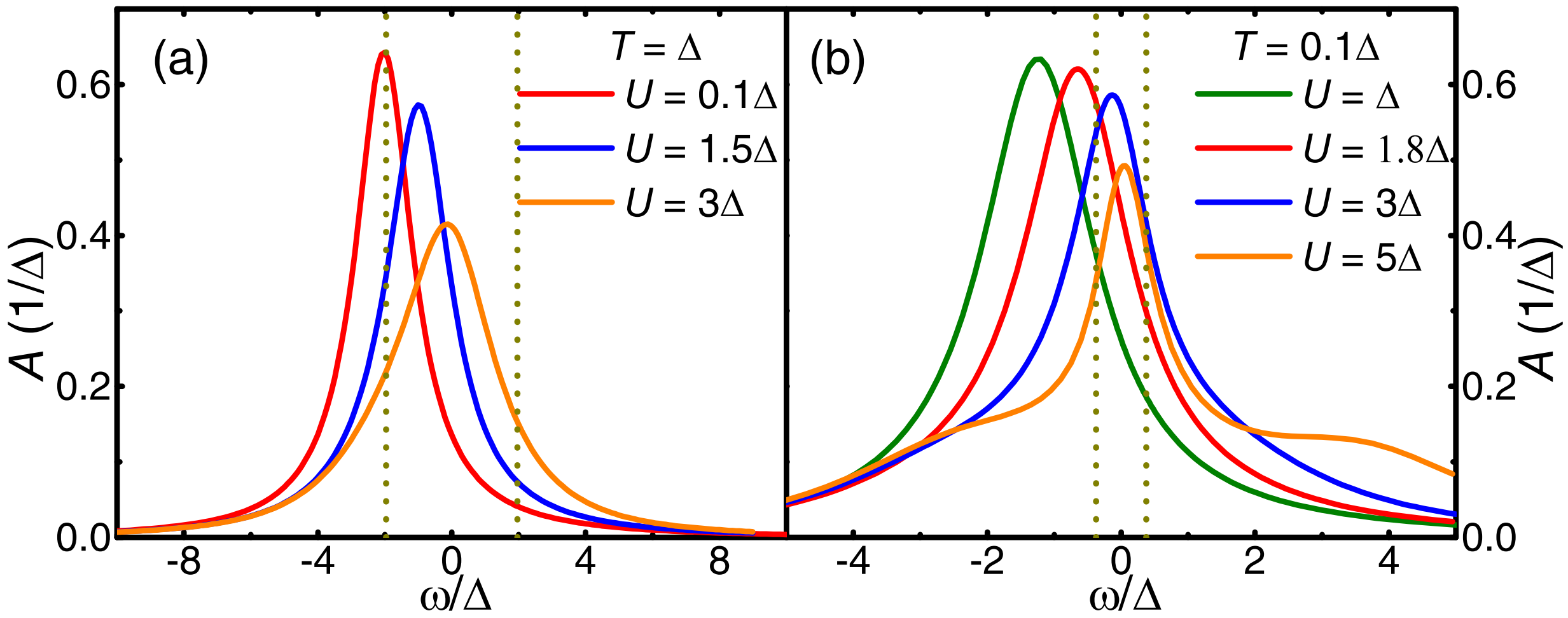

Figure S2. Dot spectral function $A(\omega)$ of a nonequilibrium QD with a varying of $U$ at the background temperature (a) $T = \Delta$ and (b) $T = 0.1\Delta$. The inset of (b) shows the Kondo spectral peak at various $T$. The vertical lines mark the excitation energy window $(\mu_L - \omega_L, \mu_R + \omega_R)$. The parameters are the same as in the caption of Fig. 3 of the main text.

## VI. DOT SPECTRAL FUNCTION IN DIFFERENT REGIMES

Figure S2 depicts the HEOM calculated dot spectral functions of interacting QDs with different values of $U$. As shown in Fig. S2(a), at a relatively higher background temperature ($T = \Delta$) the renormalized Hubbard peak gradually moves into the excitation energy window with the increase of $U$. For instance, for the QD with $U = 1.5\Delta$ the renormalized Hubbard peak resides largely within the excitation window, and hence the QD is in a resonant situation. While at a lower background temperature ($T = 0.1\Delta$), besides the evolution of Hubbard peak, Kondo resonance states start to emerge at a sufficiently large $U$. As shown in Fig. S2(b), for the QD with $U = 5\Delta$ a prominent Kondo spectral peak forms at the center of the excitation energy window. The presence of Kondo resonance states is clearly demonstrated by the inset of Fig. S2(b), in which the peak height increases continuously with the lowering of temperature.

* xz58@ustc.edu.cn
† diventra@physics.ucsd.edu